# Resonance photon yield from solid Xe below the bottom of the n=1 excitonic band

A.N. Ogurtsov,[1,2] E.V. Savchenko,[1] S. Vielhauer[3] and G. Zimmerer[3]

[1]Institute for Low Temperature Physics and Engineering of NASU, Lenin Avenue 47, 61103 Kharkov, Ukraine
[2]National Technical University "KhPI", Frunse Street 21, 61002 Kharkov, Ukraine
[3]Institut für Experimentalphysik der Universität Hamburg, Luruper Chausee 149, 22761 Hamburg, Germany

VUV-luminescence spectroscopy is a powerful tool to investigate the final step of the relaxation of electronic excitations in rare gas solids — the radiative decay of the emitting centers. The photon excitation spectra of various luminescence bands provide us with information about interplay of different channels in the course of relaxation of electronic excitations and formation of the emitting centers. It made possible to investigate recently the effects of exciton mixing, branched relaxation of electronic excitations and various inelastic radiation-induced processes in atomic cryocrystals [1]. The measurement of luminescence within the absorption band of the lowest $\Gamma(3/2)$ n=1 excitons itself is attended by inevitable coincidence of the wavelength of exciting and emitting photons and usually it was not carried out in the experimental works. At the same time the polaritonic nature of excitons in rare gas solids reveal itself just in the range at (and below) the bottom of the lowest $\Gamma(3/2)$ n=1 excitonic band. The absorption spectra of excitons in rare gas solids exhibit the pronounced low-energy tails and photon yield from the samples at selective excitation in the region of the absorption edge carries the information about interplay of the processes of photon absorption, reflection, scattering and emission in the course of interaction of incident photons with the sample [2].

Time-resolved fluorescence spectroscopy under selective photoexcitation by synchrotron radiation in the VUV allows studying in real time scale the creation of electronic excitations and their relaxation. In the present study the combination of the time-resolved spectroscopy with spectroscopy under selective photoexcitation has been used to study the photon yield from solid Xe in the energy range below the bottom of the lowest $\Gamma(3/2)$ n=1 excitonic band. The experiments were performed at the SUPERLUMI experimental station at HASYLAB, DESY, Hamburg. Selective photon excitation was performed with $\Delta\lambda$=0.2 nm. The luminescence was spectrally dispersed by 0.5 m Pouey monochromator with $\Delta\lambda$=2 nm equipped with multisphere plate detector.

Figure 1c shows the decay curves of resonance photon yield from solid Xe measured in the energy range 7.9–8.36 eV at $T$=9 K. All decay curves were measured at emission energies which were coincide with excitation energies. The numbers of curves in Fig.1c correspond to numbers at data points in Fig.1a. The feature at $t$=2.7 ns is a side-bunch. Curve 1 represents the stray light with lifetime 0.1 ns. Naturally it contributes to all decay curves. In addition, the relatively slow component appears when excitation energy growths along the low polaritonic branch in the range below the threshold $E_I$=8.18 eV [2,3] of direct photoabsorption by molecular centers [4]. Above $E_I$ the intensity of signal is too low and the points in this range are just the rough approximation of the lifetime. The energy dependence of the lifetime for decay curves (Fig.1c) is shown in Fig.1a. The general behavior of decay curves in Fig.1c is similar to correspondent behavior of decay curves of free-exciton luminescence in solid Kr [5]. To separate the contribution of free-exciton line to the total photon yield it is necessary to provide the analysis of the luminescence using high-resolution secondary monochromator.

*Acknowledgements*. The support of the DFG grant 436 UKR 113/55/0 is gratefully acknowledged.

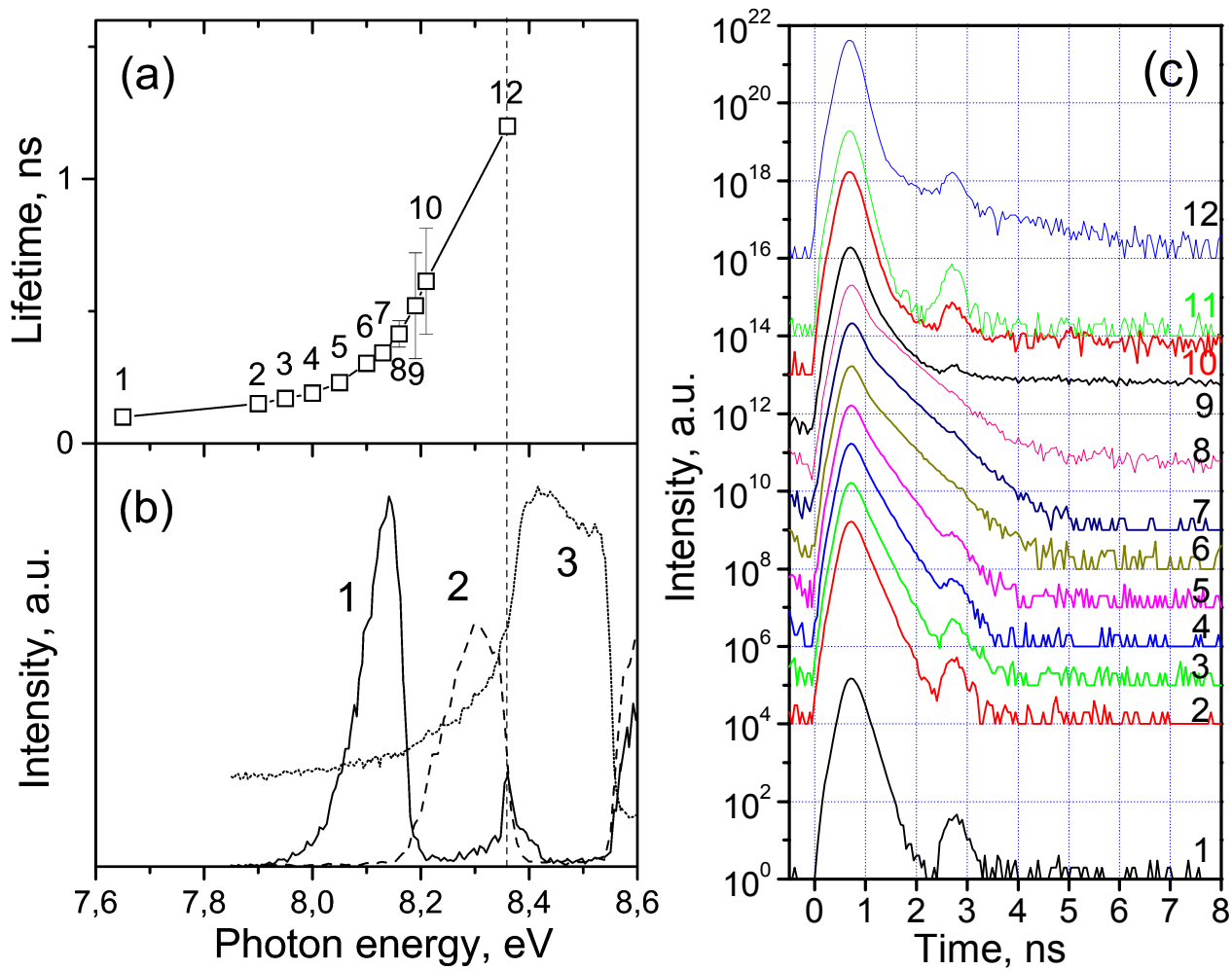

Figure 1: (a) – Lifetimes of decay curves from Fig.1c.
(b) – Excitation spectra of free-exciton (curve 1) and molecular (curve 2) bands of solid Xe, measured at their maxima; and reflection spectrum (curve 3).
(c) – Decay curves of resonant photon yield, recorded at photon energies denoted at Fig.1a. All data were measured at $T$=9 K.